\documentclass[aps,pra,epsfigure,twocolumn]{revtex4}
\usepackage{dcolumn}    
\usepackage{bm} 
\usepackage{graphicx}
\usepackage{amsmath}    
\usepackage{latexsym}
\usepackage{amsfonts}   
\usepackage{amssymb}
\usepackage{array}      
\usepackage{epsfig}
\usepackage{txfonts}
\usepackage[colorlinks=true,linkcolor=blue,urlcolor=blue,citecolor=blue]{hyperref}
\usepackage{color}
\usepackage{hyperref}

\newcommand{\ket}[1]{\left\vert#1\right\rangle}
\newcommand{\bra}[1]{\left\langle#1\right\vert}

\newcommand{\Tr}{\text{Tr}}
\renewcommand{\vec}[1]{\boldsymbol{#1}}

\begin{document}
 \title{Nonclassicality and criticality in symmetry-protected magnetic phases}
       
\author{Matthew J. M. Power}
\affiliation{Centre for Theoretical Atomic, Molecular and Optical Physics, Queen's University Belfast, Belfast BT7 1NN, United Kingdom}
\author{Steve Campbell}
\affiliation{Centre for Theoretical Atomic, Molecular and Optical Physics, Queen's University Belfast, Belfast BT7 1NN, United Kingdom}
\author{Maria Moreno-Cardoner}
\affiliation{Centre for Theoretical Atomic, Molecular and Optical Physics, Queen's University Belfast, Belfast BT7 1NN, United Kingdom}
\author{Gabriele De Chiara}
\affiliation{Centre for Theoretical Atomic, Molecular and Optical Physics, Queen's University Belfast, Belfast BT7 1NN, United Kingdom}

\begin{abstract}
Quantum and global discord in a spin-1 Heisenberg chain subject to single-ion anisotropy (uniaxial field) are studied using exact diagonalisation and the density matrix renormalisation group (DMRG). We find that these measures of quantum nonclassicality are able to detect the quantum phase transitions confining the symmetry protected Haldane phase and show critical scaling with universal exponents. Moreover, in the case of thermal states, we find that quantum discord can increase with increasing temperature.
\end{abstract}
\date{\today}
\maketitle

\section{Introduction}
The study of strongly-correlated magnetic systems has experienced a tremendous boost thanks to inputs from quantum information processing \cite{amicoRMP}. In particular, the analysis of various forms of entanglement has revealed deep connections to quantum phase transitions \cite{Osborne2002,Osterloh2002,Vidal03} and order parameters \cite{DeChiara2012,Bayat}, conformal field theory \cite{Calabrese04} and numerical simulations of quantum many-body  systems \cite{Vidal_sim, CiracPEPS}. The fidelity approach has been successful for the analysis of quantum phase transitions (QPT) with no order parameters or with infinite order \cite{Zanardi,You}.

An alternative approach based on quantum correlations has allowed the quantum information/condensed matter community to analyse the ``quantumness" of not only the ground state of magnetic Hamiltonians, but also thermal states. For the latter, entanglement tends to disappear quite rapidly with temperature and is subject to mathematical pathologies such as the ``sudden death" \cite{Zyczkowski}. 
Quantum discord (QD) is a measure of nonclassicality that has attracted a lot of attention recently~\cite{Ollivier-Zurek,Henderson,discordRMP}. While the resource nature of QD is still an open question~\cite{resourceQD}, it has shown to be a remarkably effective tool in studying a range of quantum phenomena and protocols including QPTs~\cite{QPTsDiscord,campbell}, entanglement distribution~\cite{entanglementdist}, and work extraction~\cite{campbellJPB}. While most of the literature is devoted to the discord of two-level systems and continuous variable Gaussian states, very little has been done for higher dimensional quantum states~\cite{Rossignoli}. This can be understood in light of the difficulty in evaluating QD~\cite{huang}. Thus, this work aims at providing a concrete analysis of the QD in a realizable physical system beyond those typically examined, namely a spin-1 chain.

In this paper, we study the quantum discord of spin-1 chains governed by a Heisenberg Hamiltonian and in the presence of single-ion anisotropy, i.e. a uniaxial (quadratic Zeeman) field. The model gives rise to three gapped magnetic phases (N\'eel, Haldane, Large-D) separated by a second order transition and a Gaussian one respectively. The Haldane phase, a symmetry protected phase, is quite peculiar as it is characterised by the absence of local order but by the establishment of a hidden string order parameter.

 While this model has been studied extensively in the condensed matter and quantum information communities using the entanglement and fidelity~\cite{Botet1983,Glaus1984,Schulz1986,Papa,EspostiBoschi,Chen,Campos2006,Albu2009,Pollmann2010,Rodriguez10,DeChiaraPRB2011,Hu2011,lepori,Langari2013,ejima}, a systematic investigation of its discord content is missing. The aim of this paper is to fill this gap by analysing the two-spins discord and the global version introduced in Ref.~\cite{rulli}. We employ numerical simulations based on exact diagonalisation and density matrix renormalisation group (DMRG) \cite{dmrg2,dmrg1}. We find that discord is able to locate very accurately the two transitions by showing singular behaviour as a function of the single-ion anisotropy. Moreover, for the Gaussian transition we are able to extract an estimate for the critical exponent for the correlation length. Finally, we analyse the thermal behaviour of discord and find that while it normally decays with increasing temperature, in the large-D phase and for two non nearest-neighbour spins it actually increases.
 
 The paper is organised as follows: in Sec.~\ref{sec:discord} we revise the measures of nonclassicality used in the rest of the paper; in Sec.~\ref{sec:model} we revise the magnetic properties of the spin-1 model we consider; finally in Sec.~\ref{sec:results} we show our numerical results and in Sec.~\ref{sec:conclusions} we conclude.

\section{Measures of nonclassicality}
\label{sec:discord}
The quantum discord (QD) between two systems $A$ and $B$ described by a density matrix $\rho_{AB}$ can be defined as the difference of two distinct ways to measure correlations in a quantum system that would otherwise give the same result classically~\cite{Ollivier-Zurek,Henderson}. The first way is to use the quantum mutual information
 \begin{equation}
I(\rho_{AB}) = S(\rho_A)-S(\rho_A|\rho_B),
\end{equation}
 where $S(\rho_A) = -\Tr[\rho_A\log_2\rho_A]$ is the von Neumann entropy of the reduced state $\rho_A =\Tr_B \rho_{AB} $ of system $A$ and analogously for system $B$. The quantity
 \begin{equation}
S(\rho_A|\rho_B) = S(\rho_{AB})-S(\rho_B),
\end{equation}
is the conditional entropy.
 An alternative definition of correlations can be given in terms of information acquired on $A$ after performing a measurement of $B$ with a set of projectors $\{\Pi_B^j\}$. Let us call 
  $\rho_{A|j}=(1/p_j)\Tr_B[{\Pi}^j_B\rho_{AB}]$ the state of system $A$ after outcome $j$ is obtained measuring system $B$ with probability $p_j = \Tr[\Pi_{B}^{j} \rho_{AB}]$. We thus define the one-way classical information as
  \begin{equation}
J(\rho_{AB}) = S(\rho_A)-\sum_j p_j S(\rho_{A|j}).
\end{equation}
QD is the difference of the quantum mutual information and the classical one-way information, minimized over the set of orthogonal projective measurements on $B$
  \begin{equation}
\mathcal{D}^{B\rightarrow A}(\rho_{AB})=\inf_{\{\Pi_B^j\}}[I(\rho_{AB})-J(\rho_{AB})].
\end{equation}
This definition is not symmetric under the exchange of $A$ and $B$ as the measurements are performed on system $B$ only. A symmetrized version of the QD can be obtained with a bi-local measurement $\Pi_{ij} = \Pi_A^i\otimes \Pi_B^j$ such that $\Pi(\rho_{AB}) = \sum_{ij} \Pi_{ij}\rho_{AB}\Pi_{ij}$. We define the symmetric QD
\begin{equation}
\label{eq:discordsymm}
\mathcal{D}_2(\rho_{AB})=\min_{\{\Pi\}} \left\{S(\rho_{AB}||\Pi(\rho_{AB}))
-\sum_{\alpha=A,B}S(\rho_{\alpha}||\Pi^\alpha(\rho_\alpha))\right\},
\end{equation}
where we have introduced the relative entropy:
\begin{equation}
S(\rho||\sigma)=\Tr[\rho\log_2\rho]-\Tr[\rho\log_2\sigma],
\end{equation}
which vanishes as $\sigma$ approaches $\rho$.
Eq.~\eqref{eq:discordsymm} can be interpreted as the difference between the first term, that is global on $A$ and $B$, and the second term, which is the sum of two local contributions.  Eq.~\eqref{eq:discordsymm} was shown to be generalizable to multipartite states~\cite{rulli}. For a quantum system comprising of $N$ subsystems we define the global quantum discord (GQD)
\begin{equation}
\label{GQD}
\mathcal{D}_N(\rho_{N})=\min_{\{\Pi\}}\bigg\{S\left(\rho_{N}||\Pi(\rho_{N})\right)-\sum_{\alpha=1}^N S\left(\rho_{\alpha}||\Pi^\alpha(\rho_\alpha)\right)\bigg\}.
\end{equation}
In order to evaluate Eqs.~\eqref{eq:discordsymm} and \eqref{GQD} for the spin-1 system presented in the following section, we require suitable projective measurements. In Ref.~\cite{Rossignoli} the parametrization of local orthogonal measurements for spin-1 particles was given. We use the spin-1 operators $S^{x,y,z}$ fulfilling the normal angular momentum commutation relations. For simplicity, we define the eigenstates of the $z$-component of the angular momentum as: $S_z\ket{m}=m\ket{m}$ with $m=-1,0,+1$.
A projective measurement for three-level systems is specified by three orthogonal projectors summing to the identity matrix
 \begin{equation}
\sum_{m=0,\pm1} \ket{m_A}\bra{m_A}  = \openone,
\end{equation}
where $A$ is a unitary matrix and we have defined the transformed basis states  as
\begin{equation}
 \ket{m_A} = A\ket m.
\end{equation}

Contrary to spin-1/2 systems, the norm of the Bloch vector for pure states, $\vec{P}=\langle \vec{S} \rangle$ is not always one for spin-1 systems, i.e. they are not always coherent states. This is related to the fact that, while for spin-1/2 particles unitaries can always be written as spin rotations (apart from an irrelevant phase factor),  for spin-1 there exist more general unitaries related to quadrupolar operators, which induce spin squeezing.
 Thus, the most general unitary should be written as the exponential of a polynomial of degree 2 in the spin operators. Alternatively one can split this exponential as the product of the exponentials of simpler combinations of spin operators. 

Following Ref.~\cite{Rossignoli}, we first define the states
\begin{eqnarray}
\ket{ m_r } &=& \exp\left[i\left(\gamma\left(S^{z}\right)^2 -\gamma - \phi_0 S^z\right)\right]  \times \nonumber \\
&\times& \exp\left[-i\alpha\left(S^x S^y + S^y S^x\right)\right] \times \nonumber \\
&\times& \exp\left[\frac{i\beta}{\sqrt{2}} \left(S^y +S^y S^z +S^z S^y\right)\right] \; \ket{m }.
\end{eqnarray}
Then the most general basis is obtained by rotating the states $\ket {m_r}$ in any possible direction using the following combination of rotations:
\begin{equation}
\label{eq:genrot}
\ket{m_A} = e^{-i \psi S^x} \; e^{-i \theta S^y} \; e^{-i \phi S^z} \ket{m_r}.
\end{equation}
It is therefore sufficient to parametrize the most general orthonormal basis of spin-1 systems with six coefficients (since $\phi_0$ is constrained by the other parameters~\cite{Rossignoli}).

%%%%%%%%%%%%%%%%%%%%%
\section{The model}
\label{sec:model}
We conduct our analysis on the ground and thermal states of a spin-1 chain described by the Heisenberg Hamiltonian with uniaxial anisotropy of strength $U$
\begin{eqnarray}
\label{eq:bilin}
\mathcal{H} = \sum_{i}  S^{x}_{i} S^{x}_{i+1} +S^{y}_{i} S^{y}_{i+1}  + S^{z}_{i} S^{z}_{i+1}+U \sum_{i}  \left(S^{z}_{i} \right)^2,
\end{eqnarray}
where the subscript $i$ runs over the sites in the chain and we will consider both open and periodic boundary conditions. When $U<0$ $(U>0)$ the anisotropy is usually referred to as ``easy-axis" (``easy-plane") anisotropy. 
The ground state phase diagram consists of three phases: for $U \gtrsim 0.968$ the system is in the ``large D" phase \cite{note_largeD} in which the ground state has a strong component onto the $\ket{00\dots 0}$ state. For $U< -0.315$, the system is in the N\'eel antiferromagnetic phase, characterized by the staggered magnetization. For the intermediate values of $U\!\in\![-0.315,0.968]$, the system is in the Haldane phase, a symmetry protected phase, characterized by the absence of local ordering, a nonlocal string-order parameter and an entanglement spectrum with even degeneracies \cite{Pollmann2010,lepori}. The transition separating the N\'eel and Haldane phases is of the Ising type, with the staggered magnetization ordering in the N\'eel phase. The transition from Haldane to large D is of Gaussian type and has been recently studied in \cite{Hu2011}. The entanglement properties of spin-1 chains have been studied using the block entropy and entanglement spectrum \cite{EspostiBoschi, Pollmann2010,lepori}, thus connecting with predictions from conformal field theory.

When computing discord we need to optimize the basis so as to minimize the quantities introduced in the previous section. The ground and thermal states of the real symmetric Hamiltonian $\mathcal{H}$ can be chosen real. It follows that the optimal basis is always real as in Ref.~\cite{campbell}. This further reduces the number of parameters to optimize over since we impose
$$
\gamma = \psi =\phi = \phi_0 = 0,
$$
so that the basis $\ket{m_A}$ is a real superposition of the states $\ket{m}$. We remark that this constraint is only applied to the optimization of the GQD, which due to computational complexity, necessitates such simplifications. However, when dealing with the symmetric QD, Eq.~\eqref{eq:discordsymm},  a full minimization with the full set of angles is tractable. 

 %%%%%%%%%%%%%%%%%%%%%%%
\begin{figure}[t]
\includegraphics[width=0.95\columnwidth]{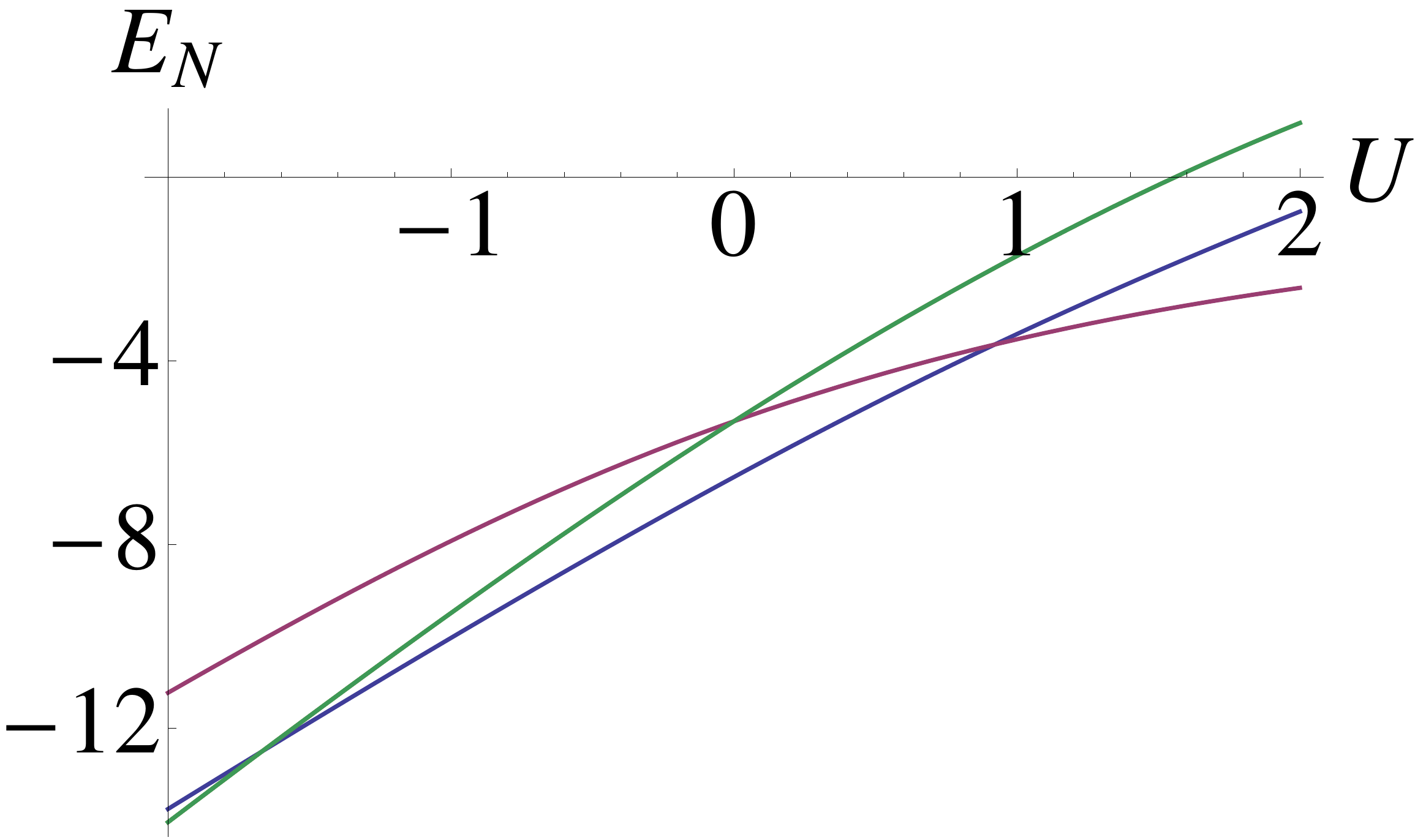}
\caption{(Color online) Lowest three energy levels of Hamiltonian Eq.~\eqref{eq:bilin} with periodic boundary conditions as a function of $U$ for $L=5$. For odd length chains the ground state changes suddenly due to energy level crossings. Here we see the first crossing at $U\simeq -1.6$ and a second crossing at $U\simeq 0.9$.}
\label{fig:energylevelgraph}
\end{figure}
%%%%%%%%%%%%%%%%%%%%%%%

\section{Nonclassicality in the Spin-1 Heisenberg Model}
\label{sec:results}
When considering finite-length chains, we find an even/odd parity effect with the total length, the origin of which lies in a geometrical frustration for odd lengths. In fact, in the repulsively interacting Heisenberg chain we are considering, nearest-neighbor spins tend to form strongly correlated pairs, an effect observed in the alternating behavior of the block entropy and other correlations. Thus, while for an even chain all spins are paired, for odd chains there is always an unpaired spin. This frustration gives rise to energy crossings as observed in Fig.~\ref{fig:energylevelgraph}. This means that, at these energy crossings, the ground state of the system changes discontinuously with $U$. For this reason, we examine even and odd lengths separately, as for the latter we will observe discontinuities in the discord measures. However, we remark that this is a finite size effect that will vanish in the thermodynamic limit, as the three phases of the model are all gapped.
%%%%%%%%%%%%%%%%%%%%
\begin{figure}[t]
\includegraphics[width=0.95\columnwidth]{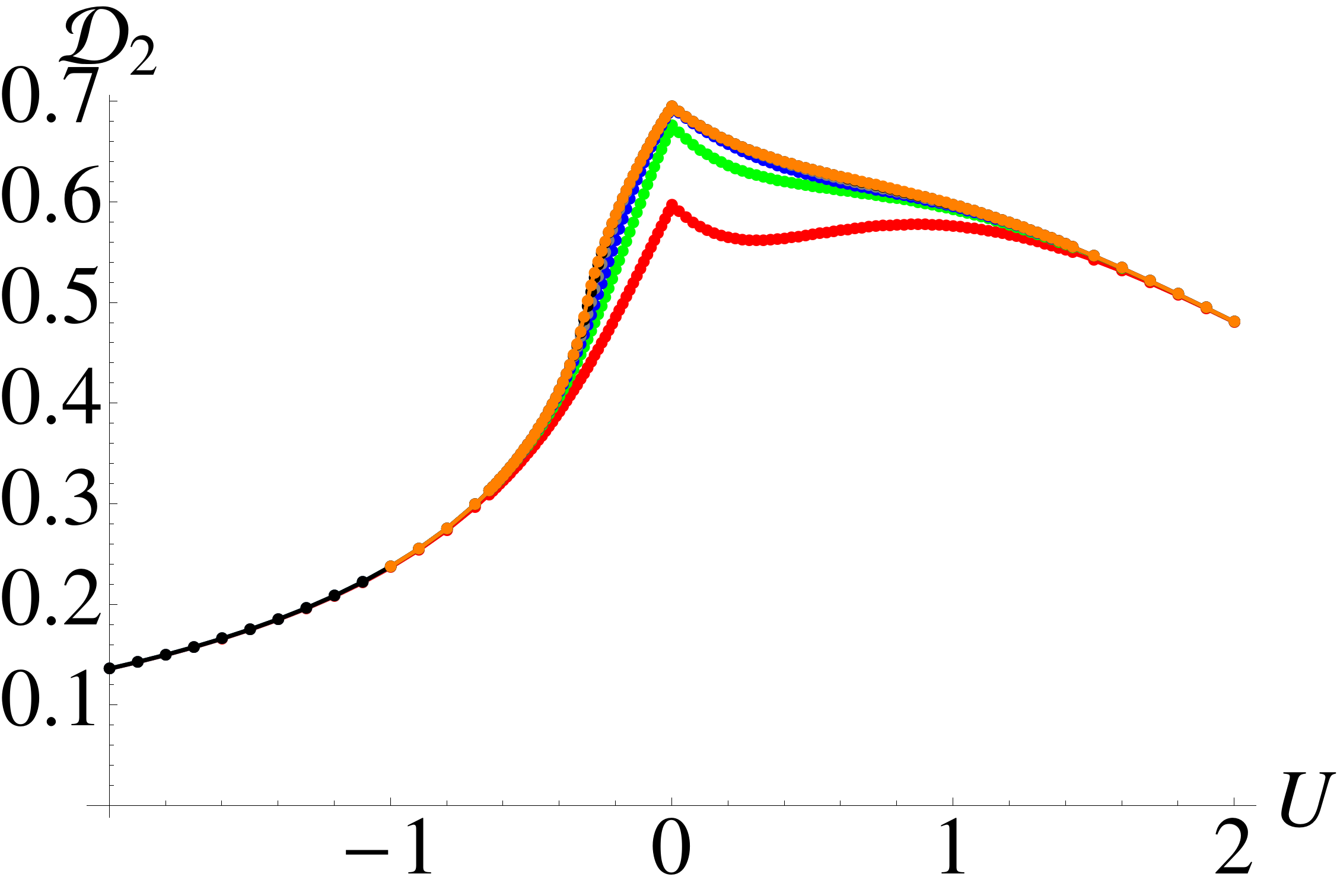}\\
\caption{(Color online) Nearest-neighbor symmetric QD, $\mathcal{D}_2$, for the reduced state of the two central spins of an open-ended chain of length $L=$ 8 (red), 16 (green), 32 (blue), 64 (gray), 128 (black) and 256 (orange) going from bottom to top. Notice that the curves for $L=32$, 64, 128, and 256 are almost indistinguishable except near $U\!\sim\!-0.3$.}
\label{DMRG}
\end{figure}
%%%%%%%%%%%%%%%%%%%%%  

\subsection{Nearest-neighbor spins}
We begin by analyzing the reduced state of the two central spins for the thermal ground state when open ended-boundary conditions are imposed on Eq.~\eqref{eq:bilin}. Further, to capture the pertinent features of the model, we consider only even-$L$ and, thus avoid pathological features due to energy level crossings in the ground state when $L$ is odd. Through DMRG calculations, we are able to determine the reduced state of the two central spins and calculate the symmetric discord Eq.~\eqref{eq:discordsymm}. In Fig.~\ref{DMRG} we plot $\mathcal{D}_2$ for $L=8$, 16, 32, 64, 128 and 256 spins. We see several interesting features emerging for increasingly large chains. For $U<-0.6$ and $U>1.6$, we see the curves for all lengths have collapsed on top of each other and the reduced states are virtually identical. This indicates that, in these regions (which are sufficiently far from the QPTs of the model), already with 8 spins, we are close to the properties of the thermodynamic limit. In the intermediate region, $U\in[-0.6,1.6]$, the curves exhibit a much richer behavior, and as we increase $L$ the value of the QD increases. 
%%%%%%%%%%%%%%%%%%%%
\begin{figure}[t]
(a)\\
\includegraphics[width=0.95\columnwidth]{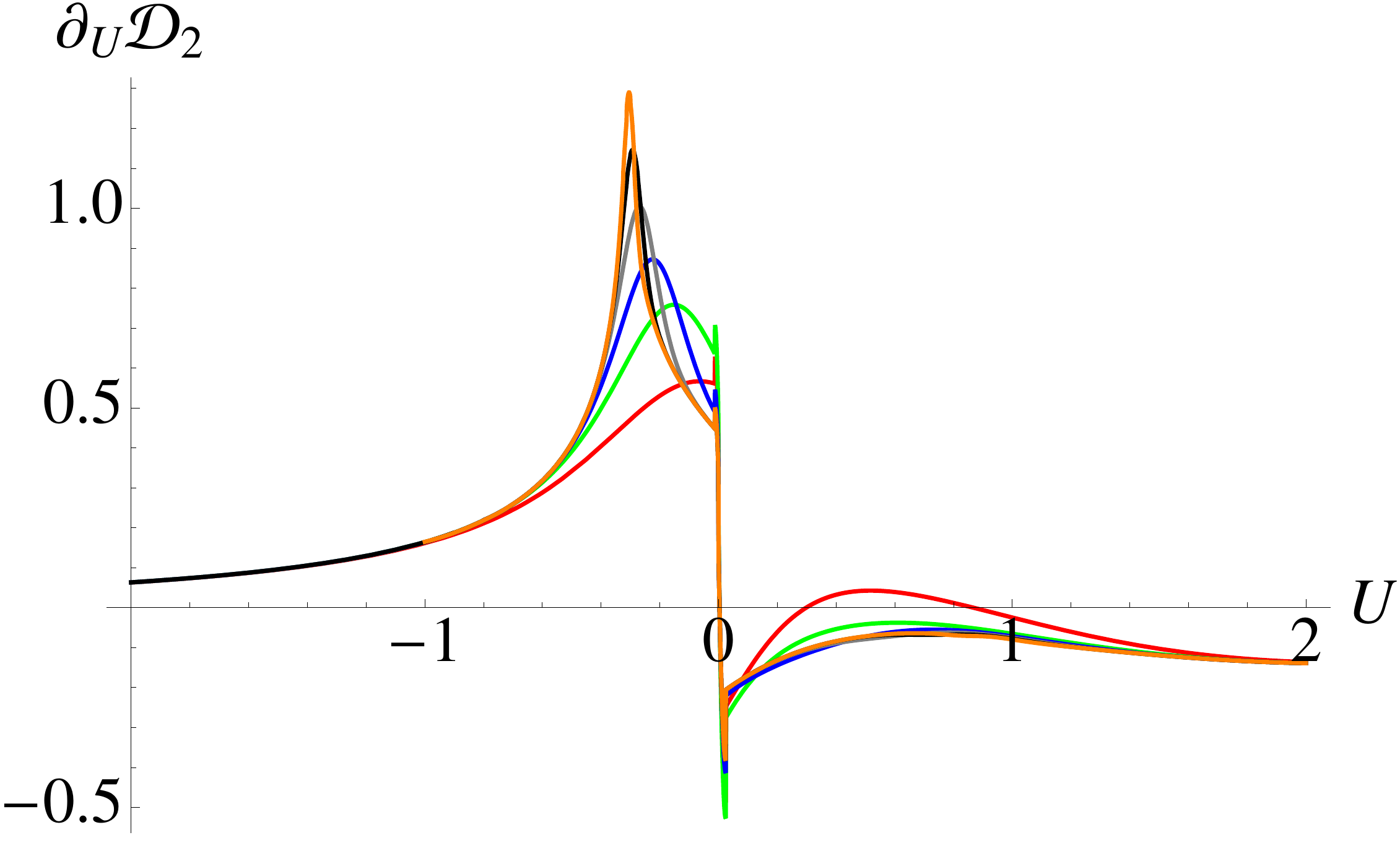}\\
(b)\\
\includegraphics[width=0.95\columnwidth]{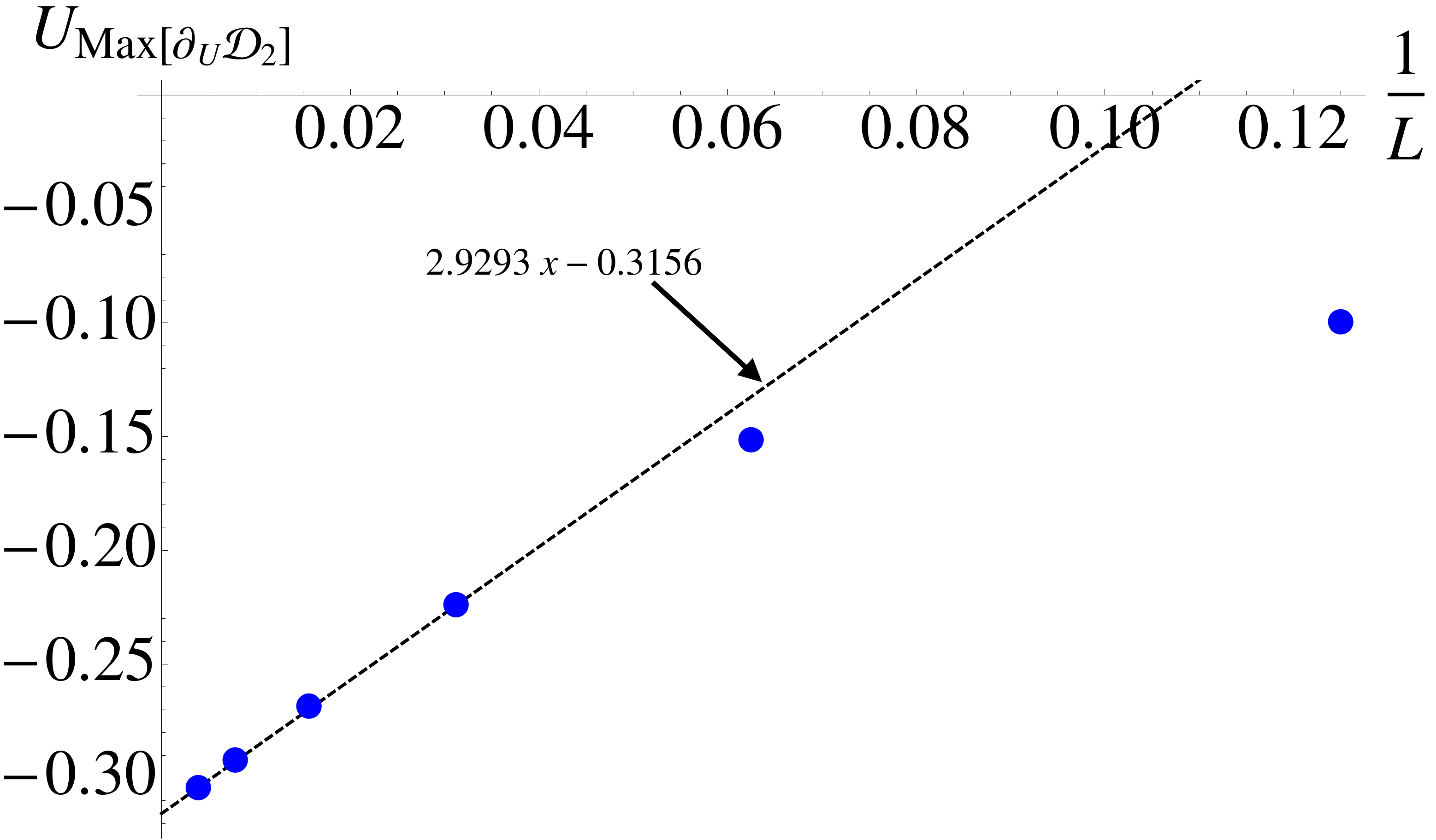}
\caption{(Color online) (a) Derivative of the nearest neighbor symmetric QD against $U$ for $L=$ 8 (red), 16 (green), 32 (blue), 64 (gray), 128 (black) and 256 (orange). (b) Finite size scaling of the value of $U$ where the derivative is maximum against inverse chain length.}
\label{DMRG2}
\end{figure}
%%%%%%%%%%%%%%%%%%%%%  
A striking feature is the cusp at $U=0$. This corresponds to the point when the optimizing angles required to minimize $\mathcal{D}_2$ change. When $U<0$ we find $\mathcal{D}_2$ is optimized when both spins are measured using the angles $\theta=\alpha=\beta=0$, corresponding to a projection onto the eigenbasis of $S_z$,  while for $U>0$ we require $\theta=\pi/2$, $\alpha=\beta=0$, corresponding to a projection onto the eigenbasis of $S_x$. For $U=0$ both sets of angles give identical values of $\mathcal{D}_2$. 
Such a sudden change is unsurprising considering that at this point we are switching from easy-axis for $U<0$ to easy-plane for $U>0$ anisotropy. Although $U=0$ is not a critical point, the Hamiltonian, and therefore its ground state, is SU(2) invariant, and  projective spin measurements differing only by a spin rotation give the same discord. Indeed, such behavior is not uncommon when dealing with nonclassicality indicators that involve complex parameter optimizations. A similar behavior was recently reported in the spin-1/2 $XY$ model when examining measures of local quantum coherence~\cite{fanchini}.
%%%%%%%%%%%%%%%%%%%%
\begin{figure}[t]
\includegraphics[width=0.95\columnwidth]{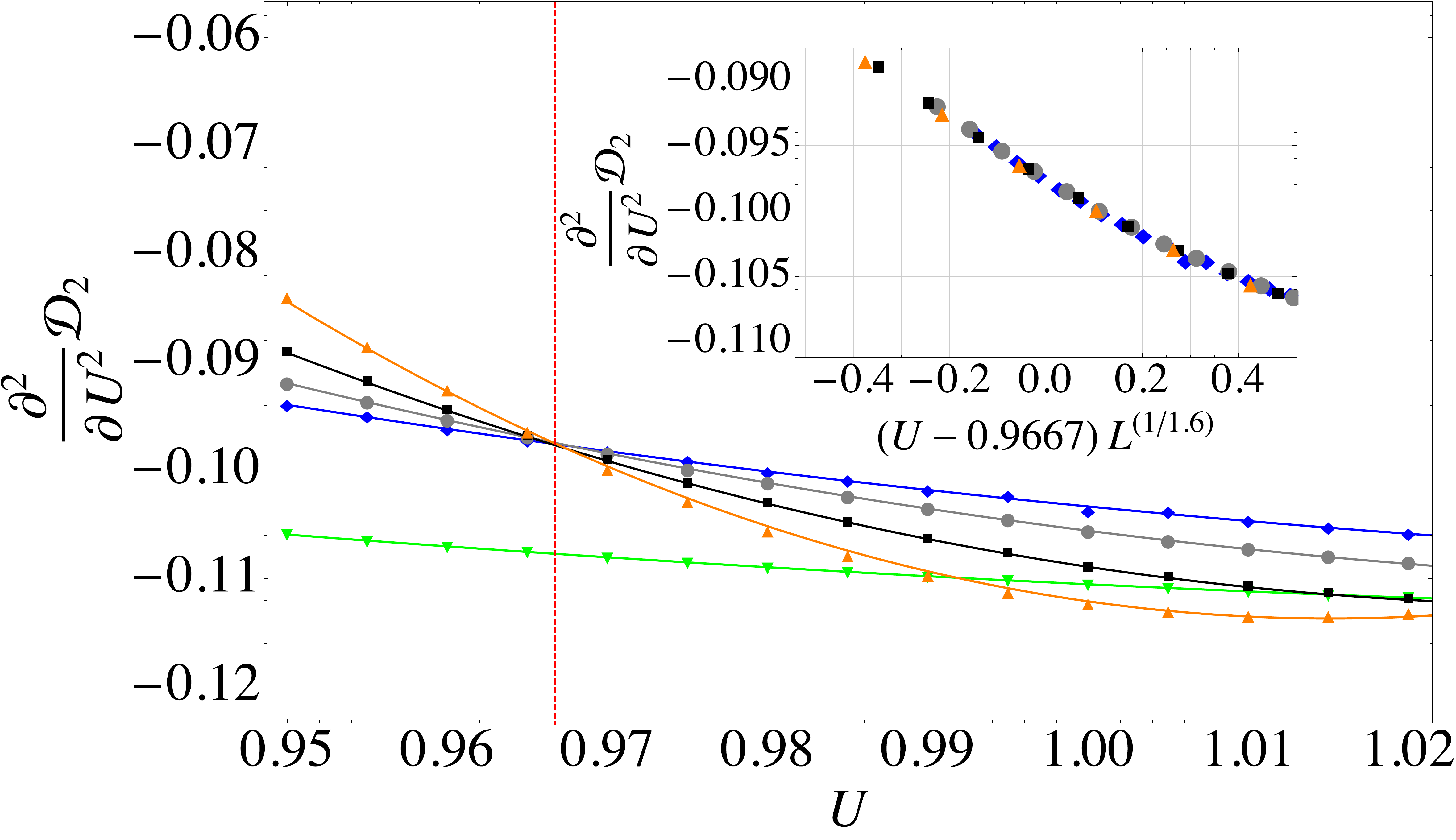}
\caption{(Color online) Second derivative of the symmetric QD with respect to $U$ in the critical region of the Gaussian QPT. The point markers are the numerically calculated values $L$=16 (downward green triangles), 32 (blue diamonds), 64 (gray circles), 128 (black squares), and 256 (upward orange triangles). The solid lines are quadratic functions of best fit for each data set. The vertical red dashed line at $U=0.9667$ is approximately where the curves cross each other. Inset: finite size scaling of the second derivative of $\mathcal D_2$ using the estimate for the critical point $U=0.9667$ and the fitted value $\nu=1.6\pm 0.1$ (see Eq.~\eqref{eq:fss}).}
\label{DMRG3}
\end{figure}
%%%%%%%%%%%%%%%%%%%%%  

In Fig.~\ref{DMRG2} (a) we examine the derivative of the QD with respect to $U$ for chains of increasing length. We see a peaked behavior appearing near $U\!\!=\!\!-0.3$, which becomes more pronounced for larger $L$. Such behavior is consistent with the signature of a second order QPT \cite{QPTsDiscord}: a discontinuity of the second order derivative of the ground state energy or of the first derivative of the state (and therefore of discord). We can accurately predict the critical value for $L\to\infty$ through finite size extrapolation. In Fig.~\ref{DMRG2} (b) we disregard the two smallest sized chains ($L=8$ and 16) and find the linear fit for the remaining four data points, which approximately lie on a straight line, giving the critical point position at $U=-0.3156$, precisely inline with the value determined in~\cite{DeChiaraPRB2011}.

Turning our attention to the Haldane-large D QPT, this transition is Gaussian and expected to be a third order transition. Furthermore, the critical region is known to be very tight, and not extending more than $\pm0.1$ from the critical point~\cite{Hu2011}. This results in its characterization being extremely difficult. In fact, in Ref.~\cite{Hu2011} the authors employed a refined DMRG technique in order to access lengths of up to 20,000 spins to determine the critical point to a high degree of accuracy, finding $U=0.96845$. In~\cite{lepori} the critical value was estimated to be $U=0.96$ by studying the finite size scaling (FSS) of the entanglement spectrum for up to $L=204$, while using Monte-Carlo simulation the predicted value was found to be $U=0.971$~\cite{Albu2009}. As the Haldane-large D transition is a third order continuous phase transition, we expect a point of inflection in the second derivative of the energy, and consequently, in the first derivative of the ground state and therefore of $\mathcal{D}_2$.
Thus, we anticipate, by examining the second derivative of $\mathcal{D}_2$, to find a minimum. Using  finite size extrapolation, as before, even with the best quadratic rather than linear fit, we find the QPT predicted at $U=0.994$ (results not shown) which is a few percent off the value predicted in Ref.~\cite{Hu2011}, indicating that the nature of this QPT will require larger sizes to accurately locate its critical point using discord. However, a curious result appears when studying the second derivative. In Fig.~\ref{DMRG3} we show the behavior of $\partial^2\mathcal{D}_2/\partial U^2$ within the critical region. The various symbols correspond to the numerically calculated values, while the solid curves are quadratic lines of best fit for each data set.  For all chains with $L>32$ we see the second derivatives cross each other near the same point located approximately at  $U=0.9667$. Smaller chains behave markedly different, although this is in keeping with the behavior of the N\'eel-Haldane transition where $L=8$ and 16 were too small to apply finite extrapolation to.  Since we expect a point of inflection in the first derivative of the discord, or equivalently, a vanishing value for third order derivative, we conjecture that close to the critical point the second derivative of the discord scales as:
\begin{equation}
\label{eq:fss}
\frac{\partial^2\mathcal{D}_2}{\partial U^2} = f[(U-0.9667)L^{1/\nu}],
\end{equation}
where $f$ is an analytic function close to $f[0]$ and $\nu$ is the critical exponent associated with the divergence of the correlation length. That is, the critical exponent associated to the second derivative of discord equals to zero, and thus, the third derivative vanishes in the thermodynamic limit. This way all the dependence on the parameter $U$ is reabsorbed in the correlation length. By fitting our results we find that the value $\nu=1.6 \pm 0.1$ collapses the data for different lengths as shown in the inset in Fig.~\ref{DMRG3}. The value we find is in agreement with the more accurate result $\nu=1.47$ found in Ref.~\cite{Hu2011}. Therefore, if our conjecture is correct, discord is not only able to locate the position of the QPT but also the universal scaling exponents associated with it.

There is an additional parity effect for these even length chains. In the above cases we consider chains such that $(L-2)/2$ is odd. Although for all even length chains the spins form nearest neighbor pairs, in this situation we are examining (the central) two spins both of which have formed a pair with their other respective nearest neighbors, and not with each other. This means the correlation between the two central spins is weaker than with their other respective neighbors. However, the larger we take $L$, the smaller this difference becomes and we see the behavior reported in Fig.~\ref{DMRG} (a), i.e. increasing QD for increasing L. In contrast, when $(L-2)/2$ is even (i.e. $L=$6, 10, 14, \dots) the two central spins correspond exactly to a dimer formed in the chain, and this results in larger values for the QD that decreases as we increase $L$. The qualitative behavior remains unaffected and, in fact, for $L>30$ any differences between these situations are negligible.

\subsection{Global measures}
While the previous section highlighted that using measures of nonclassicality applied to reduced states can capture the thermodynamic properties of systems beyond the paradigmatic spin-$1/2$ models previously studied, we now turn our attention to the global properties of finite size chains. For spin-1/2 models the GQD was shown to be remarkably effective at spotlighting the critical nature emerging from spin chains consisting of just $10$ spins~\cite{campbell}, and was able to accurately recover the correct critical exponents. Furthermore, in this paper we use the reformulation of the global discord expression found in Ref.~\cite{campbell} which considerably simplifies the optimisation of the local measurement basis. The calculation of GQD is significantly more involved than the nearest neighbor calculations performed previously and requires us to make as much use of the symmetries present in the system to simplify our calculation (in fact we remark the calculation of QD was recently shown to be NP-complete~\cite{huang}). Therefore, from here on we assume the chain has periodic boundary conditions, thus making it translationally invariant. This immediately simplifies our calculation of Eq.~\eqref{GQD} as all spins will now be optimized using the same set of angles. This is true when the ground state does not break spontaneously  translational invariance. When $L=2$ we can determine the GQD employing a full minimization and we find the angles optimizing the GQD correspond exactly to those found in the nearest neighbor calculation, $\theta=\alpha=\beta=0$ for $U\!<\!0$ and $\theta=\pi/2$, $\alpha=\beta=0$ for $U\!>\!0$. We conjecture that for a chain with periodic boundaries these will be the optimizing angles for all even and odd length chains. For $L=4,5$ we calculated the full minimization for a smaller selection of values of $U$ and confirmed the conjecture. For larger systems, we calculate the GQD using these known fixed values for the optimizing angles. Despite not being accessible to analytic proof, this approach of using numerical confirmation has proven fruitful when calculating such involved quantities~\cite{beggi}.
%%%%%%%%%%%%%%%%%%%%
 \begin{figure}[t]
 (a)\\
\includegraphics[width=0.85\columnwidth]{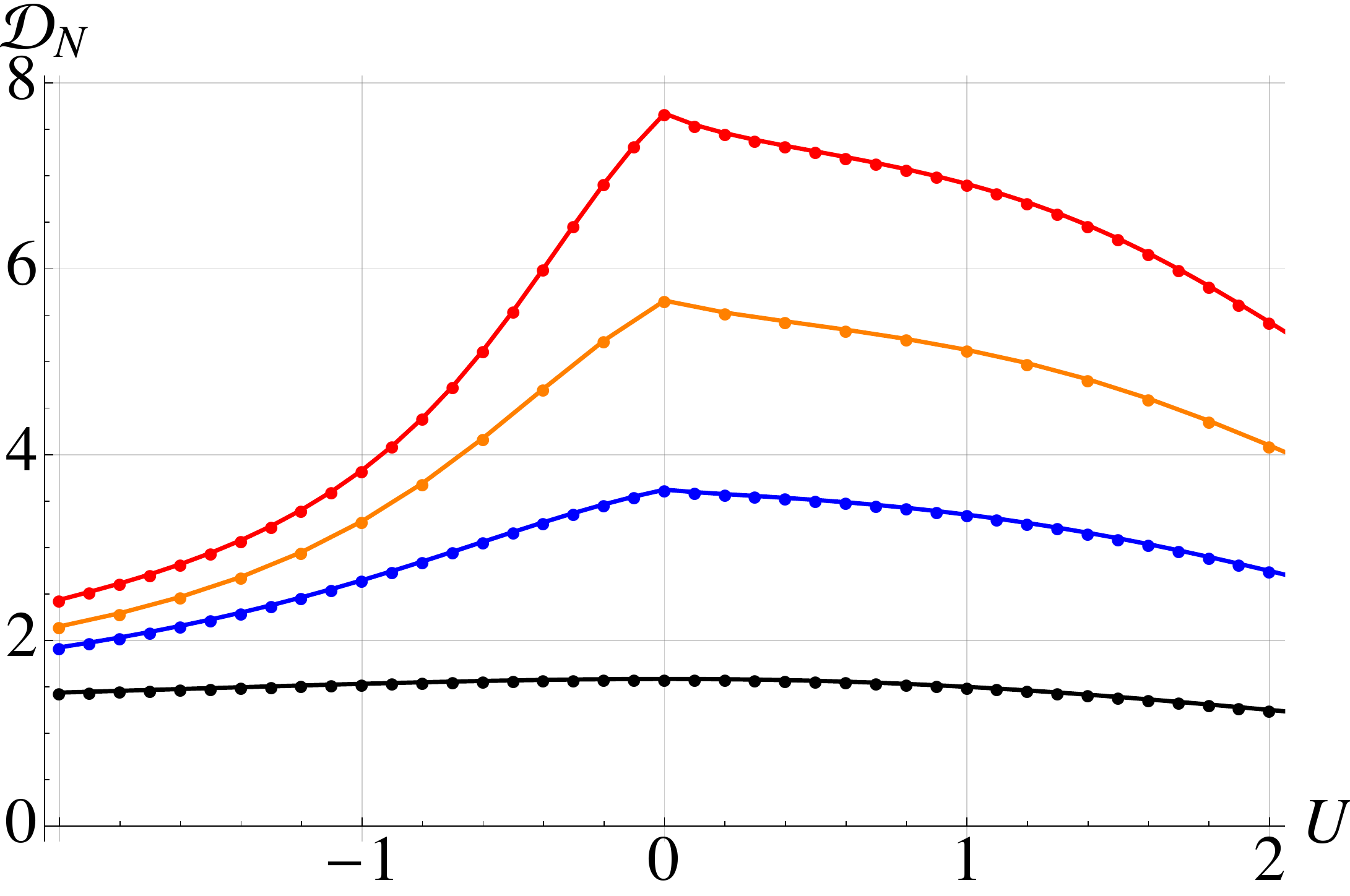}\\
(b)\\
\includegraphics[width=0.75\columnwidth]{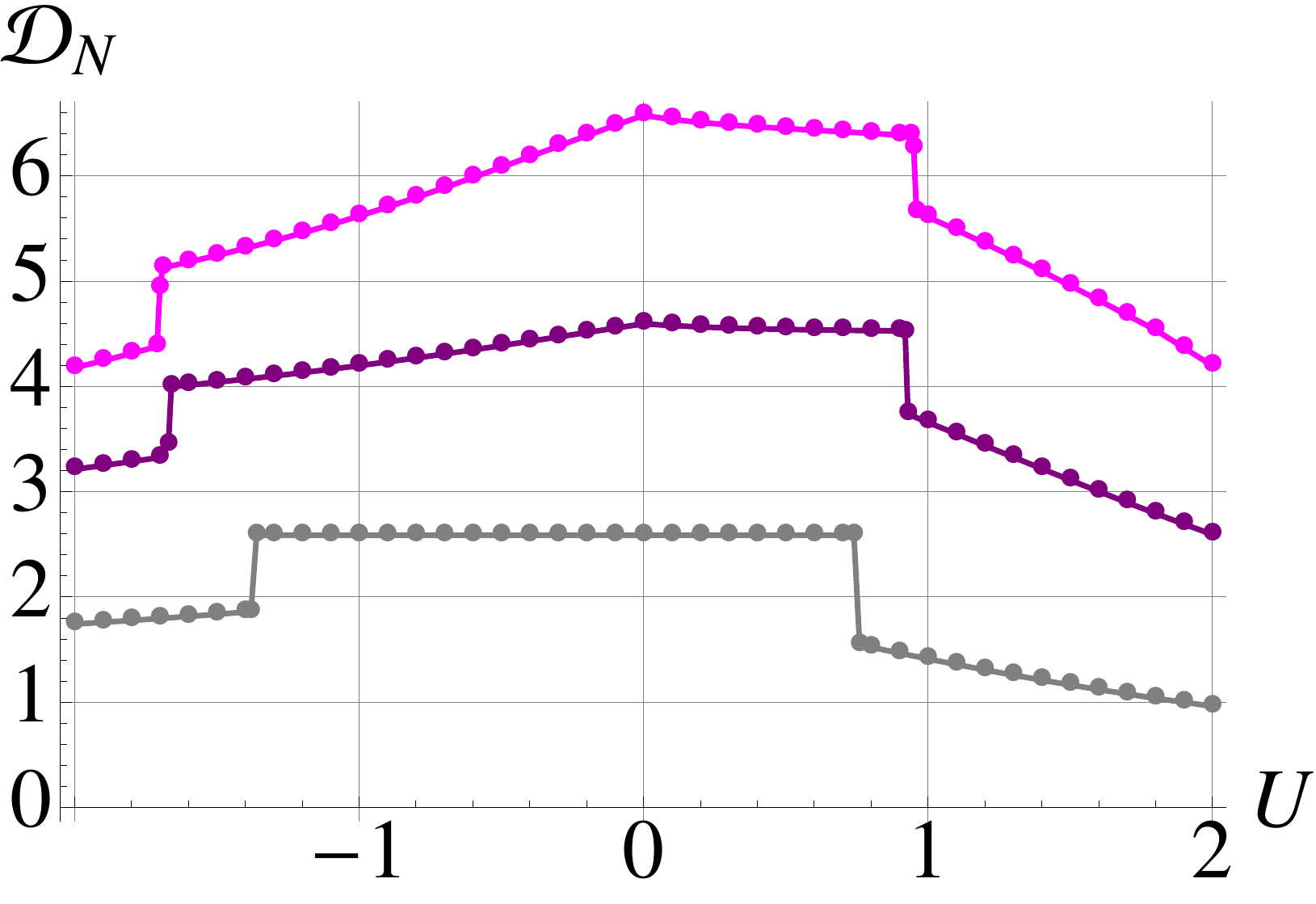}
\caption{(Color online) (a) GQD of the thermal ground state for $L=2$, 4, 6 and 8 (going from bottom to top). (b) GQD for the thermal ground state when $L=3$, 5, and 7 from bottom to top. See text for discussion. }
\label{global}
\end{figure}
%%%%%%%%%%%%%%%%%%%%%

In Fig.~\ref{global} (a) we show the GQD, Eq.~\eqref{GQD}, for finite sized chains of length $L=2, 4, 6, 8$. Consistent with the behavior of the reduced state of large chains, we see a cusp at $U=0$, which is once again a consequence of the change in the nature of the uniaxial anisotropy. We see a significant increase in the rate of change of the GQD by increasing the length. In panel (b), we examine the behavior of the GQD for odd length chains $L=3,5,7$. Here we see the sudden changes in the GQD occur when there is an energy level crossing in the ground state. For $L=5,7$ a cusp is also emerging around $U=0$. This feature is absent for the $L=3$ case whose ground state global discord is constant between the energy crossings. The sharp jumps are a consequence of the energy level crossings discussed above.

%%%%%%%%%%%%%%%%%%%%
 \begin{figure*}[th!]
 (a) \hskip0.45\columnwidth (b) \hskip0.45\columnwidth (c) \hskip0.45\columnwidth (d) \\
\includegraphics[width=0.5\columnwidth]{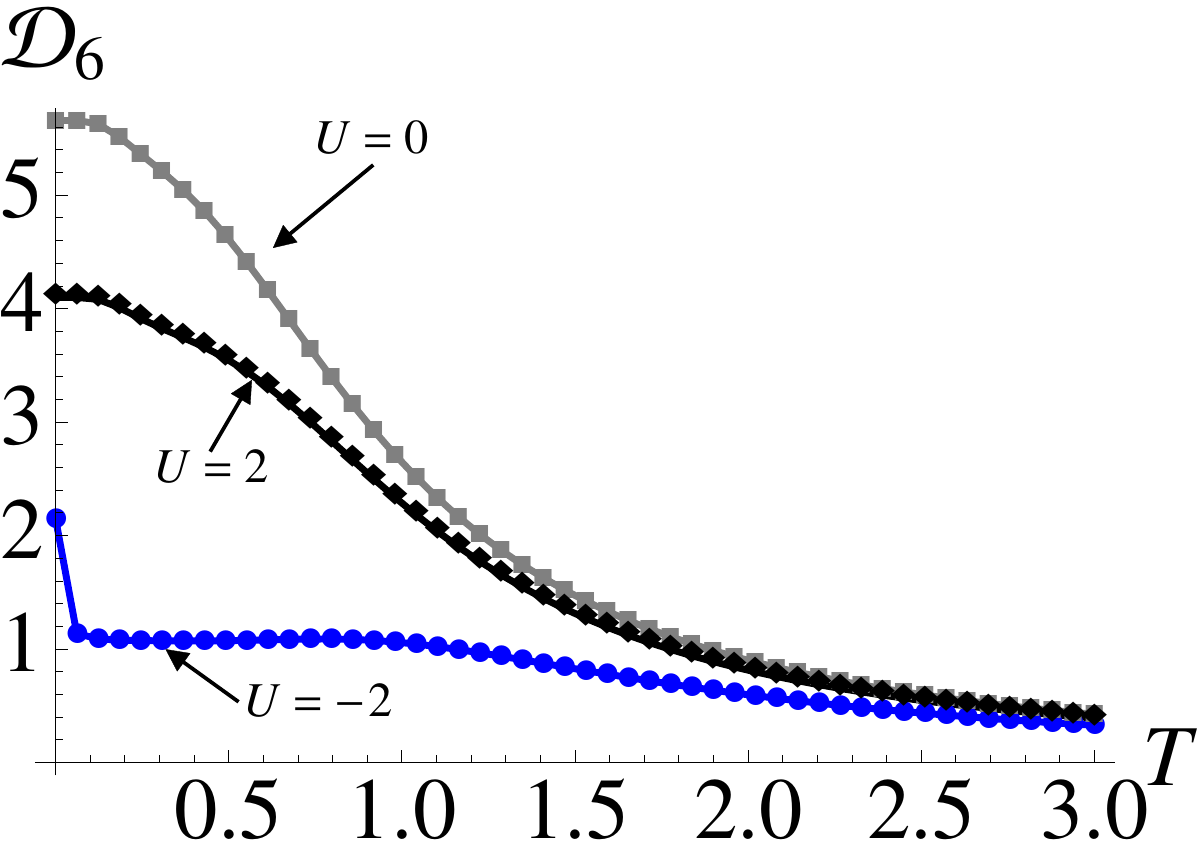}~\includegraphics[width=0.5\columnwidth]{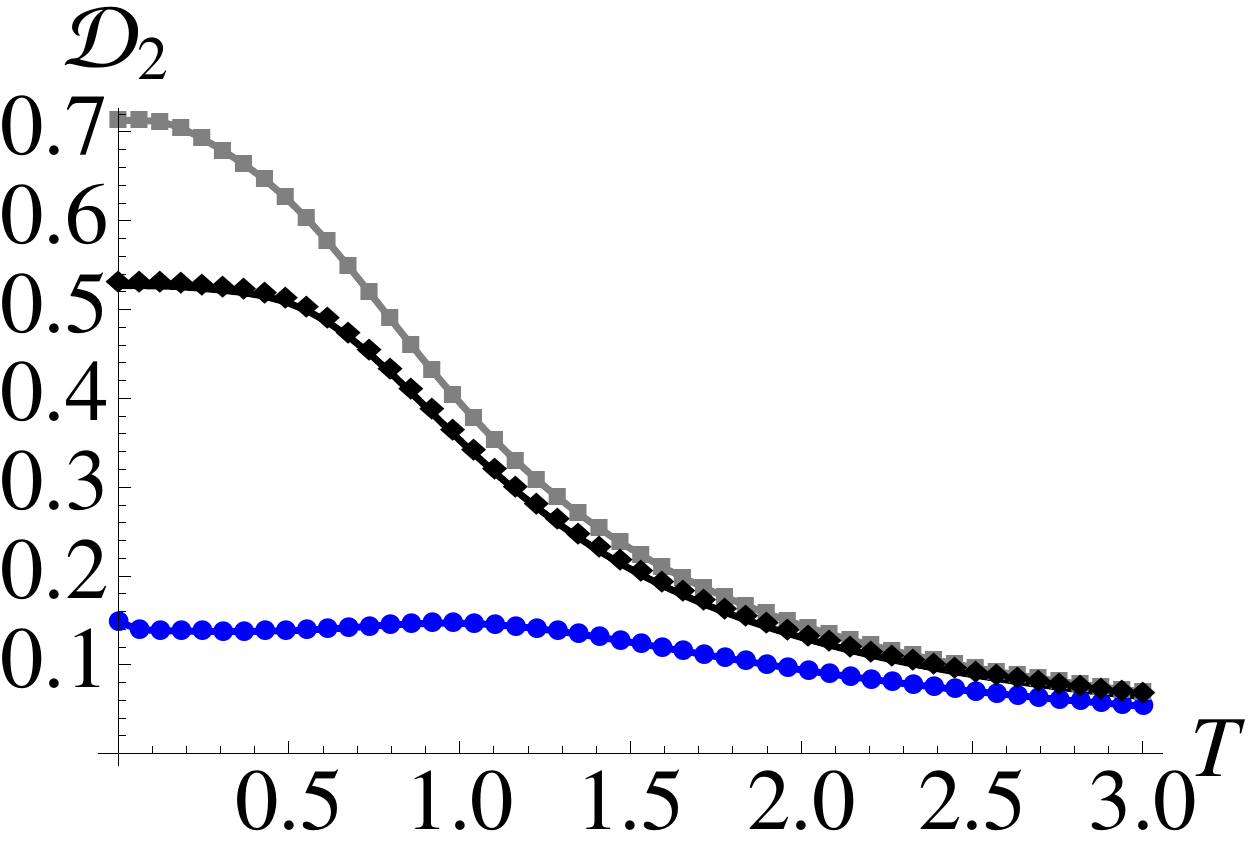}~\includegraphics[width=0.5\columnwidth]{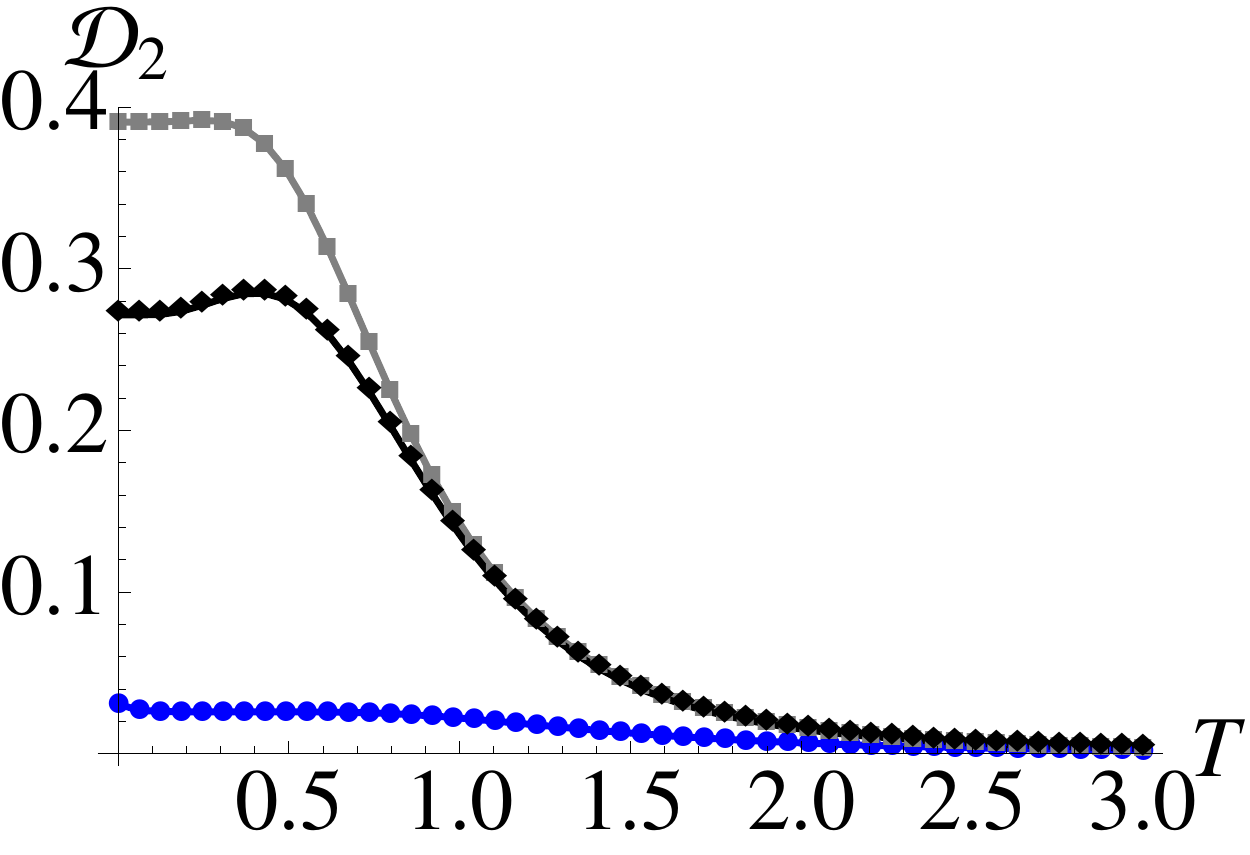}~\includegraphics[width=0.5\columnwidth]{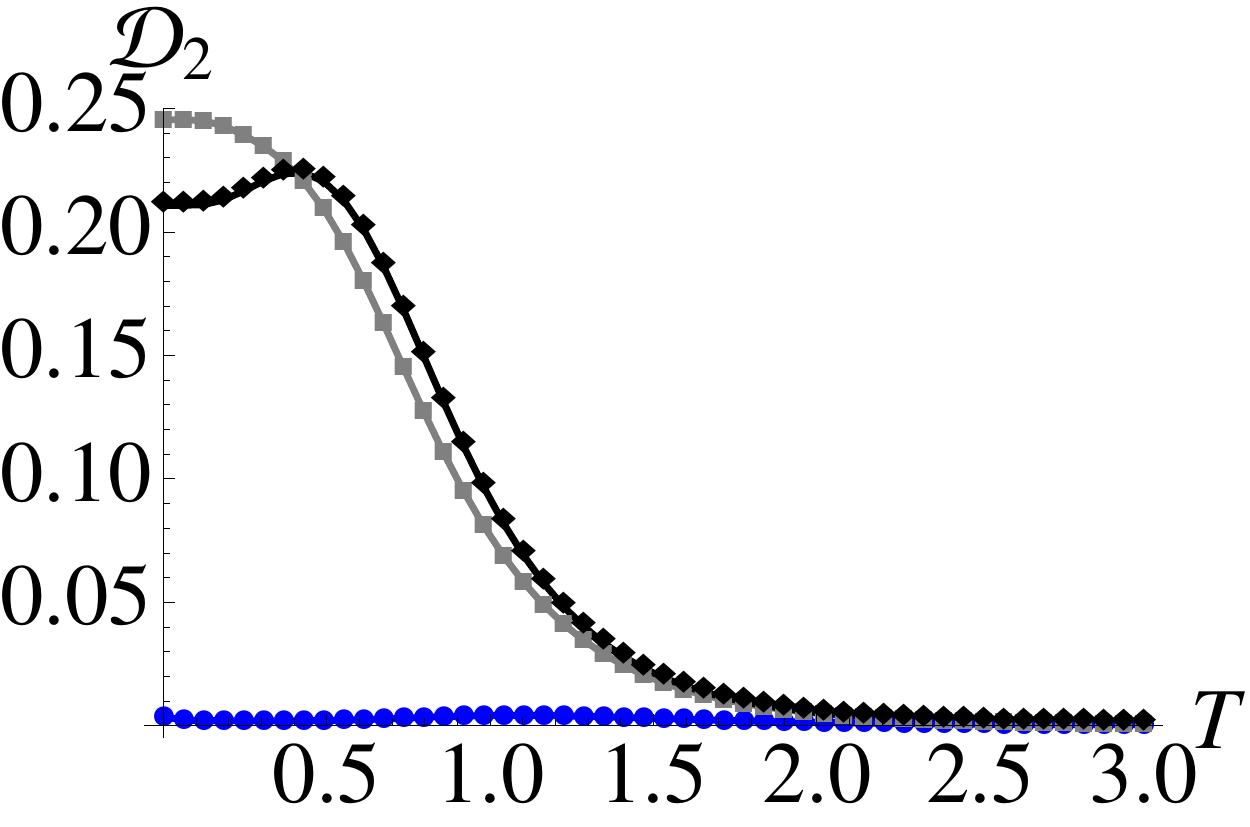}
\caption{(Color online) (a) GQD against $T$ for a closed ring of 6 spins with $U=0$ [topmost], 2 [middle curve] and -2 [bottom]. The remaining panels are the symmetric QD for the reduced state of two (b) nearest neighbor, (c) next-nearest neighbor, (d) next-next-nearest neighbor spins, from a 6 spin closed ring.}
\label{thermal}
\end{figure*}
%%%%%%%%%%%%%%%%%%%%%

A significant advantage of examining such small lengths is that it also allows us to study the thermal properties of the states. We can directly compute the thermal state as
\begin{equation}
\varrho(T)=\frac{e^{-\mathcal{H}/T}}{\text{Tr}\left[e^{-\mathcal{H}/T}\right]},
\end{equation} 
where $T$ is the temperature and we have assumed units such that Boltzmann's constant is one.
In Fig.~\ref{thermal} we fully explore the thermal effects for $L=6$ when restricted to each of the three regions of the model, although we remark that qualitatively similar results are achieved for other tractable even length chains. In panel (a), we begin with the behavior of the GQD, and it is apparent that for $U=0$ (Haldane region) and $U=2$ (large D region) the behavior of the GQD is qualitatively similar. We see a small range of $T$ where GQD is resilient to the thermal effects until $T\sim0.5$, when a quick decay occurs. In contrast, when $U=-2$ (N\'eel region), we see the magnitude of the global quantum discord is significantly less, although on the whole much more robust. However, we see a rapid initial decay that is related to a vanishingly small gap between the ground and first excited energy levels in this region. For comparison, we examine the various different two-spin reduced states, and in panel (b) we see the nearest-neighbors behave qualitatively the same as the GQD. Interestingly, for next-nearest [panel (c)] and next-next-nearest neighbors [panel (d)], the large D region shows an initial increase in ${\mathcal D}_2$ of the reduced state with increasing $T$. This unusual behavior, rarely observed for entanglement, was also noted in~\cite{james} where states of increasing mixedness can have increasing QD. In the present situation it can be explained by the presence of highly correlated excited states (similar to doblon-holon states in a Bose-Mott insulator) above a near factorised ground state in the large-D phase. A further comment is in order, as reported in~\cite{streltsov,francesco}, QD can be created by the action of local non-unital channels. While the full thermal state can be considered as the action of local channels, each such channel acts at the same rate, and therefore would appear to be incompatible with the conditions outlined in~\cite{streltsov}.

\section{Conclusions}
\label{sec:conclusions}
We have examined the nonclassical properties of the symmetry protected spin-1 Heisenberg chain with uniaxial anisotropy. Through DMRG, we were able to explore the symmetric quantum discord (QD) of the reduced state of two central spins in an open ended chain. By examining the behavior of the QD and its derivative, through finite size extrapolation, we were able to pinpoint the N\'eel-Haldane QPT in excellent agreement with the recent literature. Interestingly, we found the second derivative of the symmetric QD appeared to detect the Haldane-large D transition thanks to the third order character of this transition, and allowed us to connect its scaling with universal critical exponents. While the use of QD has been extensively applied to spin-1/2 systems, here we have shown that for higher dimensional systems the QD is still a valuable tool to explore quantum phenomena. Beyond confirming the use of bipartite QD, we further extended our analysis to global quantum discord and studied the nonclassicality of the total state, allowing us to access finite temperatures. Remarkably, we find instances in which quantum discord {\it increases} with temperature.

{\it Note:} The data used in generating all figures in this article is available from the link in Ref.~\cite{data}.

\acknowledgments
The authors thank M. Paternostro and K. Modi for useful discussions. The authors acknowledge support from the UK EPSRC (EP/L005026/1), the John Templeton Foundation (grant ID 43467), and the EU Collaborative Project TherMiQ (Grant Agreement 618074).


\begin{thebibliography}{99}
\bibitem{note_largeD} The name ``large D" phase originates from the common use of the symbol $D$ to denote the uniaxial anisotropy. As the main goal in this paper is to discuss discord which is also identified with the same letter we instead denote the uniaxial anisotropy strength with $U$ to avoid confusion.
\end{thebibliography}
  \end{document}